\pdfoutput=1 
\documentclass[12pt]{iopart}
\expandafter\let\csname equation*\endcsname\relax
\expandafter\let\csname endequation*\endcsname\relax
\usepackage{amsmath,amssymb}
\usepackage{slashed}
\usepackage{graphicx}
\usepackage{mathptmx}
\usepackage{bm}
\newcommand{\beq}{\begin{equation}}
\newcommand{\eeq}{\end{equation}}

\newcommand{\be}{\begin{eqnarray}}
\newcommand{\ee}{\end{eqnarray}}
\long\def\hidestart#1\hideend{}
\usepackage{dcolumn}
\usepackage{bm}
\usepackage{ulem}

\def\beq{\begin{equation}}

\begin{document}
\input{epsf}

\title{The consequences of $SU(3)$ colorsingletness, Polyakov Loop and $Z(3)$ symmetry  on a quark-gluon gas} 
\author{Chowdhury Aminul Islam$^a$, Raktim Abir$^a$, Munshi G. Mustafa$^{a,}$, 
Rajarshi Ray$^b$ and Sanjay K. Ghosh$^b$}
\address{$^a$Theory Division, Saha Institute of Nuclear Physics, 1/AF Bidhan Nagar, Kolkata-700 064, INDIA}
\address{$^b$Center for Astroparticle Physics and Space Science, Bose Institute, 5/EN  Bidhan Nagar, 
Kolkata-700 091, INDIA}

\begin{abstract} 
Based on quantum statistical mechanics we show that the $SU(3)$ color singlet ensemble of 
a quark-gluon gas exhibits a $Z(3)$ symmetry through the normaized character in fundamental 
representation and also becomes equivalent, within a stationary point approximation, to the ensemble 
given by Polyakov Loop. Also Polyakov Loop gauge potential is obtained by considering spatial  
gluons along with the invariant Haar measure at each space point. The probability of the normalized character in 
$SU(3)$ vis-a-vis Polyakov Loop is found to 
be maximum at a particular value exhibiting a strong color correlation. This clearly 
indicates a transition from a color correlated to uncorrelated phase or vise-versa.  
When quarks are included to the gauge fields, a metastable state appears in the temperature 
range $145\le T({\rm{MeV}}) \le 170$ due to the explicit $Z(3)$ symmetry breaking in the quark-gluon system.
Beyond $T\ge 170$ MeV the metastable state disappears and stable domains appear.  
At low temperature a dynamical recombination of ionized $Z(3)$ color charges to a color singlet  
$Z(3)$ confined phase is evident along with a confining background that originates due to
circulation of two virtual spatial gluons but with conjugate  $Z(3)$ phases in a closed loop.  
We also discuss other possible consequences of the center domains in the color deconfined phase at 
high temperature.
\end{abstract}


\date{\today}
\maketitle

\section {Introduction:}
A statistical thermodynamical description of a quantum gas, is often useful for various 
physical systems, {\it e.g.}, electrons in metal, blackbody photons 
in a heated cavity, phonons at low temperature, neutron matter 
in neutron stars, etc. In most cases the mutual interactions among the constituents
is neglected, although they interact in order to come to a thermal equilibrium.
One can imagine a situation by first allowing them to come to a thermal 
equilibrium and then slowly turning off the interactions~\cite{Gale}.  In a same perspective, 
we consider a quantum gas of non-interacting quarks ($q$), antiquarks ($\bar q$) and gluons ($g$) 
with the underlying symmetry of $SU(3)$ color gauge theory at a given temperature. Such a simple 
quantum statistical description exhibits very interesting features of a quark-gluon system.

\section{Quantum statistical mechanics and color singlet ensemble:}

In thermal equilibrium the statistical behaviour of a quantum gas
is studied through a density matrix in an appropriate ensemble as
\begin{equation}
{\rho(\beta)} =  {\rm {exp}} (-\beta
{\hat H}) \ \ , \label{denmat}
\end{equation}
where $\beta=1/T$ is the inverse of temperature and ${\hat H}$ is the
Hamiltonian of a physical system. The corresponding partition function 
for a quantum gas having a finite volume can be written as
\begin{equation}
{\cal Z} = {\rm{Tr}} \left (  e^{-\beta {\hat H}} \right ) 
= \sum_n \left \langle n \left |  e^{-\beta {\hat H}} 
\right | n \right \rangle \ \ , \label{part0} 
\end{equation}
where $|n\rangle $ is a many-particle state in the full Hilbert space 
${\cal H}$. Now, the full Hilbert space contains states which should not 
contribute to a desired configuration of the system. One can restrict
those states from contributing to the partition function by defining a 
reduced ensemble for a desired configuration as 
\begin{equation}
{\cal Z} = {\rm{Tr}} \left ( {\hat {\cal P}} e^{-\beta 
{\hat H}} \right ) = \sum_n \left \langle n \left | 
{\hat {\cal P}} e^{-\beta {\hat H}} 
\right | n \right \rangle \ \ . \label{part} 
\end{equation}
Now, let  ${\cal G}$  be a symmetry group with unitary representation 
${\hat U}(g)$ in a Hilbert space ${\cal H}$. The projection 
operator~\cite{weyl,aub,red,mus,mus2} 
${\hat {\cal P}}$ for a desired
configuration is defined as 
${\hat{\cal P}}_j = d_j \int_{\cal G} {\rm d}\mu (g) \chi^{\star}_j(g)
{\hat U}(g)$ , 
where $d_j$ and $\chi_j$ are, respectively, the dimension and the
character of the irreducible representation $j$ of ${\cal G}$ and 
${\rm d}\mu (g)$ is the invariant Haar measure. The
symmetry group associated with the color singlet configuration 
is $SU(N_c)$ and $d_j=1$ and $\chi_j=1$. 
Now the color singlet partition function for the system becomes,
\begin{equation}
{\cal Z}_S =\int_{SU(N_c)} {\rm d}\mu (g) {\rm{Tr}} 
\left ( {\hat U}(g) 
{\rm {exp}} (-\beta {\hat H}) \right ) \ \ . \label{part1} 
\end{equation}
The invariant Haar measure~\cite{weyl,aub} is expressed 
in terms of the distribution of eigenvalues of $SU(N_c)$ as
\begin{eqnarray}
\int {\rm d}\mu (g)\!\! = \!\!
 \frac{1}{N_c!}\! \left ( \prod^{N_c}_{i=1} 
\! \int\limits_{-\pi}^\pi \frac{{\rm d}\theta_l}{2\pi}\right )
\delta(\sum_i\theta_i)\!\! \prod_{i>j}\left |e^{i\theta_i}- e^{i\theta_j}\right |^2 , 
\label{measure}
\end{eqnarray}
where the square of the product of the differences of the eigenvalues
is known as the Vandermonde (VdM) determinant. The class parameter $\theta_l$ 
obeys $\sum^{N_c}_{l=1}\theta_l=0 \ ({\rm{mod}} 2\pi)$ 
ensuring  the requirement of unit determinant in $SU(N_c)$. 
This also restricts that the $SU(3)$ has only two parameter abelian subgroups 
associated with two diagonal generators, which would completely characterize 
the ${\hat U}(g)$.

Now, the Hilbert space ${\cal H}$ of a composite system has a structure 
of a tensor product of the individual Fock spaces as 
${\cal H} = {\cal H}_q \otimes {\cal H}_{\bar q} \otimes {\cal H}_g$.
The partition function  in (\ref{part1}) decomposes~\cite{aub,red,mus,mus2} in 
respective Fock spaces as
\begin{eqnarray}
{\cal Z}_S & =&
\int_{SU(N_c)}  
 {\rm d}\mu (g) \, {\rm{Tr}}  \left ( {\hat U}_q 
 e^{-\beta {\hat H}_q} \right )   
 {\rm{Tr}}  \left ( {\hat U}_{\bar q}
 e^{-\beta {\hat H}_{\bar q}} \right ) 
\, {\rm{Tr}}  \left ( {\hat U}_g
 e^{-\beta {\hat H}_g} \right ),  \label{part2}
\end{eqnarray}
where the various  $U_i(g)$ act as link variables that link, respectively, the quarks, 
antiquarks and spatial gluons in a given state of the physical system.
In each Fock space there exists a basis that diagonalizes both operators
as long as ${\hat H}_i$ and ${\hat U}_i$ commute.  
Performing the traces in (\ref{part2}) using the standard procedure~\cite{Gale,aub,mus}, 
the partition function in Hilbert space becomes
\begin{eqnarray}
{\cal Z}_S\!\! &=&\!\!\! \int_{SU(N_c)}
\!\!\!\! \!\! \!\!\! {\rm d}\mu (g) \ \ e^{\Theta_q + \Theta_{\bar q} + \Theta_g}
=\int_{SU(N_c)}{\hspace*{-0.15in}}\!\!\! {\rm d}\mu (g) \ \ e^{\Theta_p }
\ \ ;
 \label{part3}
\end{eqnarray}
\begin{eqnarray}
\Theta_p 
&=& \Theta_q + \Theta_{\bar q} + \Theta_g \nonumber \\ 
&=&
2 N_f\sum_\alpha {\rm{tr}_c}  \ln \left (1+R_q
e^{-\beta (\epsilon_{q}^\alpha -\mu_q)}
\right ) 
+ 2N_f \sum_\alpha {\rm{tr}_c}  
\ln \left (1+R_{\bar q}
e^{-\beta (\epsilon_q^\alpha +\mu_{ q})}
\right ) 
\nonumber \\
&& 
- 2\sum_\alpha {\rm{tr}_c}  
\ln \left (1-R_g
e^{-\beta \epsilon_g^\alpha}\right )
, \label{theta}
\end{eqnarray}
where $\epsilon_i^\alpha= \sqrt{(p^\alpha_i)^2+m^2_i}$. Also the quark flavor ($N_f$), their spin 
and the chemical potential $\mu$, and the polarization of gluons are introduced. The finite dimensional diagonal matrix 
$R_{q ({\bar q})}$ in the basis of the color space represents the image~\cite{weyl,aub} of the group 
element in the irreducible representation of $SU(N_c)$ as
\begin{eqnarray}
{R}_q
&=& \mbox{diag}\left(
e^{i\theta_1}\,, e^{i\theta_2}\,, e^{i\theta_3} \right)\, ;  \hspace*{0.3in}
{R}_{\bar q}
= R^\dagger_q ,   
\label{image_f}
\end{eqnarray}
with their respective characters
\begin{eqnarray}
 \chi_f=\mbox{tr}_c R_q=\sum_{i=1}^{N_c}e^{i\theta_i}; \hspace*{0.3in}
\chi_f^\dagger=\mbox{tr}_c R_q^\dagger=\sum_{i=1}^{N_c}e^{-i\theta_i}. \
\end{eqnarray}
Similarly, the character in adjoint representation is obtained as
\begin{eqnarray}
 \chi_{\mbox{adj}} &=& \chi_f\chi_f^\dagger-1=\mbox{tr}_cR_g=
\mbox{tr}_c \Big[ \mbox{diag}\big (1,1,e^{i(\theta_1-\theta_2)},
e^{-i(\theta_1-\theta_2)}, \nonumber \\
&& e^{i(2\theta_1+\theta_2)},e^{-i(2\theta_1+\theta_2)},e^{i(\theta_1+2\theta_2)},
 e^{-i(\theta_1+2\theta_2)}\Big)\Big]
\end{eqnarray}

We also define normalized characters by the respective dimension of the fundamental and 
adjoint representations as
\begin{eqnarray}
\Phi= \frac{1}{N_c} {\rm{tr}_c} R_q \,\, ,
\quad \quad 
\bar{\Phi} =\frac{1}{N_c}{\rm{tr}_c} R_q^\dagger\,\, ,
\quad \quad (N_c^2-1) \Phi_A &=& N_c^2\Phi{\bar \Phi}-1 \,.
\label{eq2}
\end{eqnarray}
As we will see in the next sec. that the magnitude of the normalized character, $|\Phi|$, in fundamental representation 
is related to the thermal expectation value of the Polyakov Loop (PL).

The VdM term in (\ref{measure}) can now be written in terms of $\Phi$ and ${\bar \Phi}$ as
\begin{eqnarray}
 \prod_{i>j}^3 \left | e^{i\theta_i} - e^{i\theta_j} \right |^2 
&=& 27 [1-6\Phi {\bar \Phi} +4 (\Phi^3+{\bar \Phi}^3) 
-3(\Phi{\bar \Phi})^2 ]
= 27 \ H(\Phi,{\bar \Phi}) \, , \label{vdm}
\end{eqnarray}
for $SU(3)$. Further, this is in general not possible for $N_c>3$ as there are more than two 
independent parameters.
Now the Jacobian for variable transformation from $\{\theta_1,\theta_2\}$ to $\{\Phi , {\bar \Phi}\}$
can be obtained as
\begin{eqnarray}
 J(\Phi,{\bar \Phi})  = {(1/9)} \sqrt{27 H(\Phi,{\bar \Phi})}.
\end{eqnarray}
In the infinite volume $V$, one also needs to replace the discrete single particle sum by an integral as
$ \sum_\alpha \rightarrow (V/(2\pi)^3 \int d^3p \ .$

After performing the color trace of the matter and gauge parts, and expressing in terms of 
the characters of the fundamental and its conjugate, Eq.({\ref{part3}}) becomes
\begin{eqnarray}
{\cal Z}_S&=& \int_{SU(3)} \ d\Phi \ d{\bar {\Phi}} \ e^{\Theta_q+\Theta_{\bar q}+\Theta_g+\Theta_H}; \nonumber \\
\Theta
&=&\Theta_{q} + \Theta_{\bar q}+\Theta_g +\Theta_H \nonumber \\ 
&=& 2VN_f\int \frac{d^3p}{(2\pi)^3} 
\ln \Big [ 1 + e^{-3\beta\epsilon_q^{ +}} +  
N_c\left( \Phi + \bar{\Phi}e^{-\beta\epsilon_q^{+}}\right)e^{-\beta\epsilon_q^{ +}}
\Big] \nonumber \\
&& + 2VN_f\int \frac{d^3p}{(2\pi)^3}  
  \ln \left [ 1 +   e^{-3\beta\epsilon_q^{-}}
+ N_c\left( {\bar \Phi} + {\Phi}e^{-\beta\epsilon_q^{-}} \right) e^{-\beta\epsilon_q^{-}} \right] \nonumber \\
&& -2V\int \frac{d^3p}{(2\pi)^3}   
\ln\Big( 1 + \sum_{m=1}^8 a_m\, e^{-m\beta\epsilon_g}\Big ) +\frac{n}{2}\ln H  
.  \label{theta1}
\end{eqnarray}
with $\epsilon^\pm_q={\epsilon_q\mp\mu}$ and the coefficients $a_m$  are given by
\begin{eqnarray}\label{eq11}
a_1 &=& a_7
= 1 - N_c^2\bar{\Phi}\Phi\,,
\quad a_8
= 1\, , 
\nonumber\\
a_2
&=&
a_6
= 1 - 3 N_c^2 \bar{\Phi}\Phi
{}+ N_c^3 \left( \bar{\Phi}^3 + \Phi^3\right)\,,
\nonumber\\
a_4
&=&
2\left[
-1 + N_c^2 \bar{\Phi}\Phi - N_c^3\left( \bar{\Phi}^3 + \Phi^3\right)
{}+ N_c^4 \left( \bar{\Phi}\Phi \right)^2 \right ] \, , 
\nonumber\\
a_3
&=&
a_5
= -2 + 3 N_c^2 \bar{\Phi}\Phi
{}- N_c^4\left( \bar{\Phi}\Phi \right)^2\, .
\end{eqnarray}
We note here that the square of the VdM determinant in (\ref{measure}) enters at each point in space in the action. 
So, in the partition function the logarithm of the product  of the VdM term  should be proportional to $n$ as 
$\Theta_H = n\ln \sqrt H$, assuming $n$ is the number of points in the space. Also a constant normalization factor 
is dropped as it is subleading.

Now performing the integrations using the method of stationary points, one can write
\begin{eqnarray}
 {\cal Z}_S^0 \!\! &=& \!\! \left. e^{\Theta_q+\Theta_{\bar q}+\Theta_g +\Theta_{H}}
 \right |_{\Phi \rightarrow \Phi_0 \atop {\bar \Phi}\rightarrow {\bar \Phi_0} }
, \label{part_ic}
\end{eqnarray}
where the stationary values of $\Phi_0$ and 
${\bar \Phi_0}$ can be obtained from the extremum conditions as
\begin{eqnarray}
\frac{\partial \Theta}{\partial \Phi } &=&0 \, ; \, \, \,\,\,\, 
 \frac{\partial \Theta}{\partial {\bar \Phi} }  = 0\, . \label{extremum}
\end{eqnarray}
The color singlet thermodynamic potential density at the stationary points in infinite 
volume limit becomes
\begin{eqnarray}
{\Omega}_S^0  &=& -\frac{T}{V}\ln {\cal Z}_S^0 = 
-\frac{T}{V}\left[ \Theta_q+\Theta_{\bar q}+\Theta_g+\Theta_{\mbox{H}} \right ]_{\Phi \rightarrow 
\Phi_0 \atop {\bar \Phi} \rightarrow {\bar \Phi_0}}
 \nonumber \\
&=& \Big [ \Omega_q+\Omega_{\bar q}+\Omega_g
-\kappa T \ln H \Big ]_{\Phi \rightarrow 
\Phi_0 \atop {\bar \Phi} \rightarrow {\bar \Phi_0}} 
 =  {\Omega}^{\mbox{PL}} (\Phi_0,{\bar \Phi_0}). 
\label{potl} 
\end{eqnarray}
where $`$PL' stands for Polyakov Loop~\cite{polyakov,pisarski,pnjl}. This exhibits that the $SU(3)$ color singlet ensemble 
of a quark-gluon gas is equivalent to that of the Polyakov Loop.  We derived the ensemble of the Polyakov Loop model 
phenomenologically by imposing $SU(N_c)$ color singlet restriction on a quark-gluon gas. Also the Polyakov Loop gauge potential 
is obtained here as $\Omega_g-\kappa T\ln H$, by considering the spatial gluons in the ensemble. In some other context the gluons 
as quasiparticles in Polyakov Loop model was also studied~\cite{mislinger,sasa} but in most phenomenological studies the form of 
the gauge potential has usually been used~\cite{pnjl,bass} by fitting pure gauge lattice data. 

We note here that the calculation begun with a discussion of color 
neutrality effects in the free quark-gluon gas~\cite{aub,red,mus,mus2}. These works were concerned with global color neutrality
and the effect of which becomes irrelevant for a free gas in the limit where volume goes to infinity. Here, the extension to local 
neutrality is considered using Haar measure at every spatial point to address the effect of confinement. Nevertheless, the well-known 
calculations of the effective potential for PL to one loop in perturbation theory~\cite{pisarski} show that the local VdM contribution  
due to the Haar measure is canceled out when spatially longitudinal gluon fields ($A_0(t,{\mathbf x})$) are integrated over. 
This becomes a problem to use local Haar measure as the basis for a fundamental theory of confinement and presently it is not known yet how
to use it. On the other hand, allowing PL field where $A_0(t,{\mathbf x})$ is constant (see in the next sec.) as ${\mathbf x}\rightarrow \infty$, 
the VdM term contributes to the partition function in (\ref{theta1}) and thus in the potential in (\ref{potl}). We further note that the 
contribution  from the VdM term survives infinite volume limit since the constant, $\kappa=n/2V$, is the  ratio of two large numbers leading 
to a finite value which has to be determined from the lattice equation of state of pure $SU(3)$ gauge theory. This value is found as 
$\kappa\sim 0.0075$ GeV$^3$ in the literature~\cite{bass,pnjl}. Now, the potential in (\ref{potl}) has been considered~\cite{pnjl} as a
starting point for phenomenological models of quark-gluon thermodynamics and is also coupled to  chiral model~\cite{njl,kunihiro} 
to study extensively the deconfinement and chiral dynamics together~\cite{pnjl}. 

Now, in the next part we do not intend to study thermodynamics, instead we discuss below some of the phenomenological consequences 
of $SU(3)$ colorsingletness vis-a-vis Polyakov Loop that may lead to the formation of center domains~\cite{polyakov,domain} in quark-gluon 
plasma produced in relativistic heavy-ion collisions.

\section{Polyakov Loop and normalised character in $SU(3)$:}
The Wilson line in $SU(N_c)$  in the direction of Euclidean time $\tau$ is defined~\cite{polyakov} 
as a path-ordering by
\begin{eqnarray}
W(x) = {\cal T} \exp[ig\int \limits_0^\beta A_0(x,\tau)\ d\tau] \, ,
\end{eqnarray}
where $A_0=A_0^{a}\lambda_a$ is a temporal gauge field with Gell-Mann matrices
$\lambda_a$ ($a=1,\cdots N_c^2-1$), $g$ is the strong coupling
and $\cal T$ is the path-ordering in Euclidean time. The normalized trace of the 
Wilson line with respect to fundamental representation is
known as Polyakov Loop,
\begin{eqnarray}
 L(x) =\frac{1}{N_c} {\mbox{tr}_c}W(x),  
\end{eqnarray}
where the Polyakov Loop is a complex and transforms under the global $Z(N_c)$ symmetry 
as a field with charge one as $L \rightarrow e^{{2\pi i}/{N_c}}L$.
It also acts as an order parameter for pure gauge theory since the free energy of
the heavy static quark, $F_Q$ is related to the thermal average of 
the Polyakov Loop as $|\langle L \rangle |=\exp[-\beta F_Q(T)]$. 
 
Given the role of an order parameter for 
pure gauge~\cite{polyakov}, if $|\langle L\rangle | =0$ the $Z(N_c$) is unbroken and there is no ionization of $Z(N_c)$ 
charge, which is the confined phase below a certain temperature. At high temperature the symmetry is 
spontaneously broken,  $|\langle L \rangle |\ne 0$ corresponds to a deconfined phase of 
gluonic plasma and there are $N_c$ different equilibrium states distinguished by the phase $2\pi j/N_c$ 
with $j=0,\cdots (N_c-1)$. 

In most of the phenomenological studies~\cite{pnjl} $A_0$ is assumed to be a static temporal gauge field within 
Polyakov Loop gauge\footnote{Diagonal in eigenvalues of $SU(N_c)$ group in terms of class parameter $\theta_i$. 
Since $\theta_i$ obeys $\sum_i^{N_c}\theta_i=0({\mbox{mod}}2\pi)$ ensuring that only $(N_c-1)$ parameters subgroups associated 
with two diagonal generators. For $SU(3)$  only two parameters $\theta_1$ and $\theta_2$ are sufficient to describe the
Polyakov Loop matrix.}.  $A_0$ would completely be characterized by two diagonal generators~\cite{sasa} as 
$A_0=A_0^3\lambda_3+A_0^8\lambda_8$.  
The thermal expectation of the gauge invariant Polyakov Loop  in terms of the characters 
normalized by the dimension of the fundamental representation for $SU(3)$is given in (\ref{eq2}) as
\begin{eqnarray}
 |\Phi|
= \langle L \rangle \! =\! \left \langle \frac{1}{3} {\mbox{tr}_c}
\left ({\cal T} \exp\left [ig \!\! \int \limits_0^\beta \! A_0\ d\tau\right ]\right)\right \rangle.
\end{eqnarray}
Now, assuming a gaussian approximation one can write~\cite{polyakov, bass}
\begin{eqnarray}
 |\Phi|&=& \left|\frac{1}{3} {\rm{tr}_c} R_q \right |= \sqrt{\Phi {\bar {\Phi}}}
= |\langle L \rangle |\approx
\exp\left [-\frac{g^2}{2T^2}{\mbox{tr}}_c  \langle A_0^2\rangle \right ]. \label{thfluc}
\end{eqnarray}
The dynamics of the magnitude of the normalized character, $|\Phi|$,  in fundamental representation is governed by the thermal average
of the square of the static temporal gauge field $A_0$. In the color confined phase $|\Phi|=0$ and the background temporal gauge field 
fluctuates with high amplitudes whereas in the color deconfined phase $|\Phi|=1$ and the fluctuations of the gauge field 
almost disappears. This background gauge field in the form of Polyakov Loop also interacts nonpertubatively with the quarks and gluons in 
the thermal medium as given in (\ref{potl}). 

\section{Normalised character, center symmetry and  consequences}
\begin{figure}
\begin{center}
\includegraphics[width=0.6\linewidth,height=0.5\linewidth, angle=0]{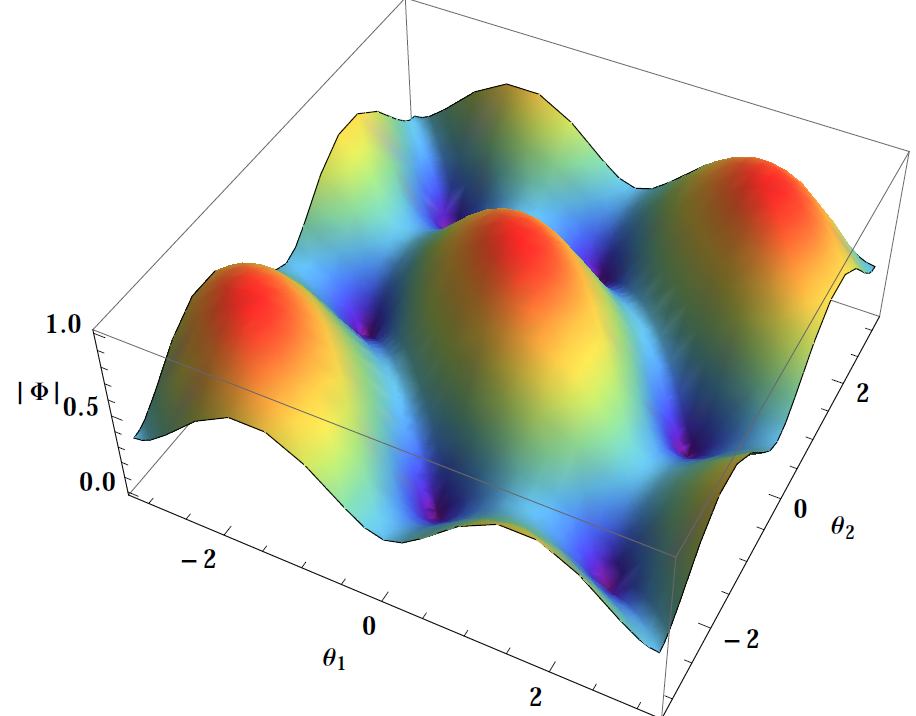}
\caption{(color online) A $3$D-plot of $|\Phi(\theta_1,\theta_2)|$, i.e.,
the normalized character in the fundamental representation of $SU(3)$ as given 
in \eqref{eq2} within $-\pi \leq \theta_1,\,\theta_2\leq \pi$.}
\label{phi_3d_f}
\end{center}
\end{figure}

\begin{figure}
\begin{center}
\includegraphics[width=0.55\linewidth,height=0.45\linewidth, angle=0]{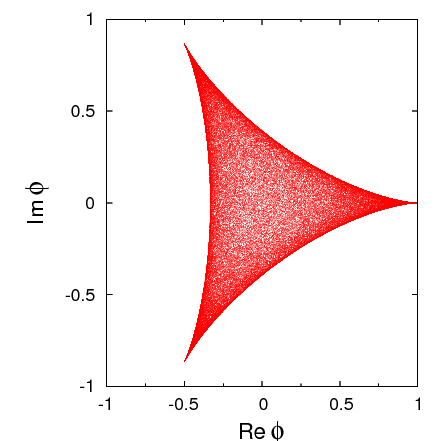}
\caption{(color online) A Monte Carlo simulation  of complex $\Phi(\theta_1,\theta_2)$
in Argand plane for which $\theta_1$ and $\theta_2$ are chosen randomly
in the domain $-\pi \le \theta \le \pi$.} 
\label{phi_comp_f}
\end{center}
\end{figure}

\begin{figure}
\begin{center}
\includegraphics[width=0.5\linewidth,height=0.4\linewidth, angle=0]{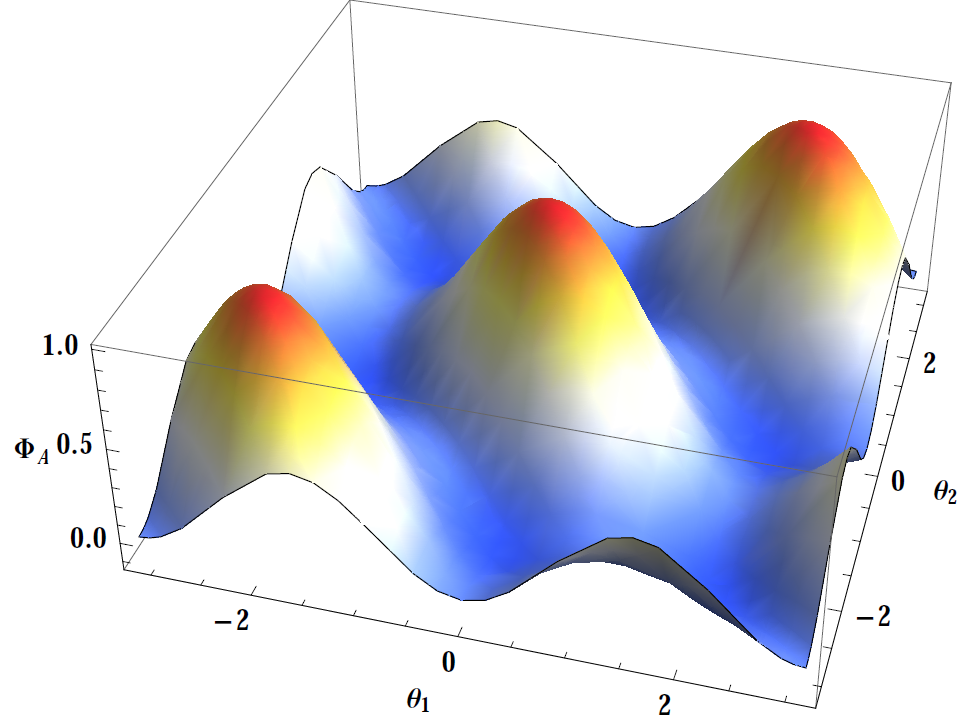}
\caption{(color online) Same as Fig.~\ref{phi_3d_f} but for   $\Phi_A(\theta_1,\theta_2)$.}
\label{phi_3d_a}
\end{center}
\end{figure}

In Fig.~\ref{phi_3d_f} a three dimensional view of 
the magnitude of the normalized character in fundamental 
representation of $SU(3)$,  $|\Phi(\theta_1,\theta_2)|$,
is shown within the domain $-\pi \leq \theta_1,\theta_2 \leq \pi$. 
It has three maxima for  $(\theta_1,\theta_2)= (0,0), \, (2\pi/3,2\pi/3) \, (-2\pi/3, -2\pi/3)$. 
There are also three minima (and three mirror images exist if
one interchanges $\theta_1\leftrightarrow\theta_2$) for $(\theta_1,\theta_2)= (0,2\pi/3), 
\, (0,-2\pi/3) \,{\mbox{and}} \, (2\pi/3, -2\pi/3)$. 
$|\Phi| $ has a
three fold degeneracy connected by a rotation of $2\pi/3$ in both 
$\theta_1$ and $\theta_2$. A Monte Carlo simulation of complex $\Phi$
is also displayed in Fig.~\ref{phi_comp_f} in a Argand plane with
 $-\pi \leq (\theta_1,\,\theta_2) \leq \pi$. 
This shows a three pointed star in a circle of unit radius, in which each point 
can be rotated by a phase $2\pi/3$ except the origin. This clearly indicates that $\Phi$  
in fundamental representation of $SU(3)$ has a center symmetry, $Z(3)$
with three rotational angles (viz., $0,\, 2\pi/3,\, 4\pi/3$ or $-2\pi/3$ ). This can also
be understood from the invariant Haar measure expressed in terms of $\Phi$ in (\ref{vdm}). Now, the 
three minima in Fig.~\ref{phi_3d_f}
uniquely correspond to the center of the circle at $\Phi=0$ in Fig.~\ref{phi_comp_f}, which
is $Z(3)$ symmetric phase or confined phase at low $T$. On the other hand, the three maxima in 
Fig.~\ref{phi_3d_f} correspond to the three pointed tips in Fig.~\ref{phi_comp_f} representing
the spontaneously broken phase or deconfined phase of $Z(3)$ at very high $T$. $\Phi$ can act as 
an order parameter for deconfinement phase transition.
In Fig.~\ref{phi_3d_a} a three dimensional plot of $\Phi_A$ is also displayed that exhibits
same features as Fig.~\ref{phi_3d_f} except minima appear in negative values, which could   
be understood from (\ref{eq2}). This clearly suggests that the magnitude of the normalized 
character in the fundamental representation of $SU(3)$ exhibits the center symmetry, $Z(3)$. 

\begin{figure}
\begin{center}
\includegraphics[width=0.5\linewidth,height=0.4\linewidth, angle=0]{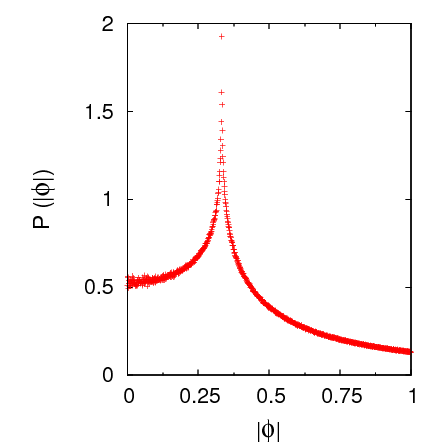}
\caption{(color online) A Monte Carlo simulation, $P(\Phi)$, corresponds to
the occurrence probability of $|\Phi (\theta_1,\theta_2)|$ in $SU(3)$ parameter
space $(\theta_1,\, \theta_2)$.
The values of $\theta_1$ and $\theta_2$ are chosen randomly within the domain 
$-\pi\leq \theta_1,\, \theta_2\leq \pi$ and then the obtained value of $|\Phi|$ 
is mapped in a given $\Phi$ bin, which is then normalized by the area of the bin.
}
\label{phi_prob}
\end{center}
\end{figure}

\begin{figure}
\begin{center}
\includegraphics[width=0.5\linewidth,height=0.45\linewidth, angle=0]{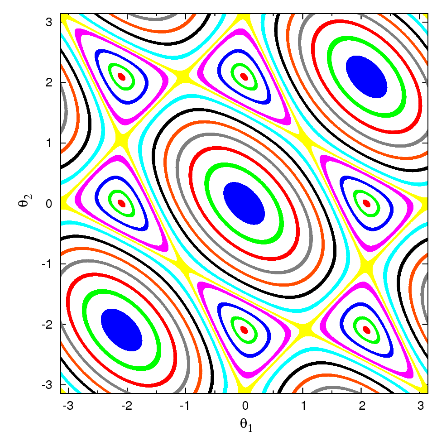}
\caption{(color online) A 2D projection of $|\Phi|$ in $\theta_1$ and $\theta_2$ plane in which 
each color corresponds to a equivalued $|\Phi|$. The red dots to purple triangles correspond to
$|\Phi|\sim0,\, 0.1,\, 0.2, \, 0.3$ whereas sea-blue lines to blue dots correspond to $|\Phi|\sim 0.4,\, 0.5,\,
0.6,\, 0.7,\, 0.8,\, 0.9, 1$. The equivalued mess connected by yellow triangles corresponds to $|\Phi|\sim 1/3$.}
\label{phi_eqva}
\end{center}
\end{figure}

\begin{figure}
\begin{center}
\includegraphics[width=0.55\linewidth,height=0.45\linewidth, angle=0]{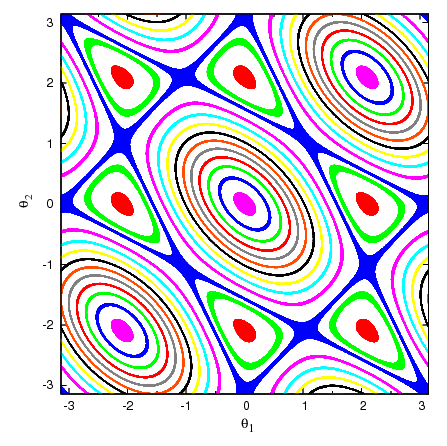}
\caption{(color online) Same as Fig.~\ref{phi_eqva} but for $\Phi_A$. 
The equivalued mess connected by blue triangles is for $\Phi_A=0$. From the blue mess to purple blobs, 
$\Phi_A$ increases by a step of $0.1$. The red dots are for $\Phi_A \sim -1/8$ whereas 
the green triangles $\Phi_A \sim -0.05$.}
\label{phi_a_eqva}
\end{center}
\end{figure}

We also noticed some more interesting features of $\Phi$ in ($\theta_1, \, \theta_2$)-plane.
In Fig.~\ref{phi_prob} a Monte Carlo simulation of the occurrence probability, $P(|\Phi|)$, 
of $|\Phi(\theta_1,\theta_2)|$ is displayed in $SU(3)$ parameter space. This plot
indicates that the maximum probability  for $|\Phi|$ to occur  when $|\Phi|=1/3$
indicating a phase transition from a color confining phase to a color deconfining phase, which has also been 
observed in Lattice QCD calculation~\cite{cheng}. This could be 
better viewed from Fig.~\ref{phi_eqva} which is a contour plot corresponding to Fig.~\ref{phi_3d_f}. 
As $T$ increases, $\Phi$ increases~\cite{pnjl} from zero in the confined phase and reaches unity for ideal gas. 
In the domain $ 0 < |\Phi| <1/3$, the color neutral states start decomposing
but prefer to reside in $Z(3)$ minima and its mirror images minima. Color charges (partons) with thermal momentum
in this domain cannot overcome the barriers provided by the large amplitude of the thermal fluctuations of the background 
gauge field in (\ref{thfluc}). This domain of $|\Phi|$ is shown by red dots ($|\Phi|\sim 0$) to purple triangles 
($|\Phi|\sim 0.3$) in Fig.~\ref{phi_eqva}. As long as such states are inside the domain of $Z(3)$ minima, a strong  color 
correlation exists among the color charges like a liquid~\cite{bass}, because the mean free path of the color charges is
of the order of size of the domain in $Z(3)$ minima. In this $|\Phi|$ domain, the normalized 
character in adjoint representation varies as $-1/8\leq \Phi_A < 0$ which is represented
in Fig.~\ref{phi_a_eqva}.

Now for $|\Phi|=1/3$, the $Z(3)$ minima disappear and get connected to each other in $Z(3)$ space, which 
is represented by the yellow mess in Fig.~\ref{phi_eqva}. This causes $P(|\Phi|)$ to be 
maximum in Fig.~\ref{phi_prob} exhibiting a long range color correlation and the thermal fluctuations 
of the background gauge field attain a critical value as the separating barriers of minima become flat and wider.  
Here $\Phi_A=0$ as is also represented by the blue mess in Fig.~\ref{phi_a_eqva}.

When $1/3 < |\Phi| \leq 1$, the correlated $Z(3)$ domains start to get uncorrelated  and 
the ionization of $Z(3)$ color charges begin which is evident from equivalued $|\Phi|$ lines in Fig.~\ref{phi_eqva} starting 
from sea-blue ($|\Phi|\sim0.4$) to green lines ($|\Phi|\sim 0.9$) at a step of $0.1$. 
When $|\Phi|\sim 1$  a complete ionization of $Z(3)$ charges take place and they reside at those maxima in Fig.~\ref{phi_3d_f} which 
are also represented by blue blobs in Fig.~\ref{phi_eqva}. This ionization can also be seen in Fig.~\ref{phi_a_eqva} 
through purple equivalued $\Phi_A$ lines to purple blobs in the range $0.1\leq \Phi_A \leq 1$.  So, in the color 
deconfined phase ($1/3 < |\Phi| \leq 1$), there are formation of domains which are also separated by the nonperturbative 
interaction of the background gauge fields. These domains of ionized color charges can act as scattering centers 
in the deconfined phase lead to jet quenching.  A hard jet after losing energy through gluon emission by 
the scattering with those ionized domain of color charges~\cite{bass} in deconfined phase ($1/3 < |\Phi| \leq 1$) 
can enter the confining phase ($0< |\Phi| \leq 1/3$) and hadronize by recombination~\cite{mus2,hadr}.

\begin{figure}
\begin{center}
\includegraphics[width=0.45\linewidth,height=0.3\linewidth, angle=0]{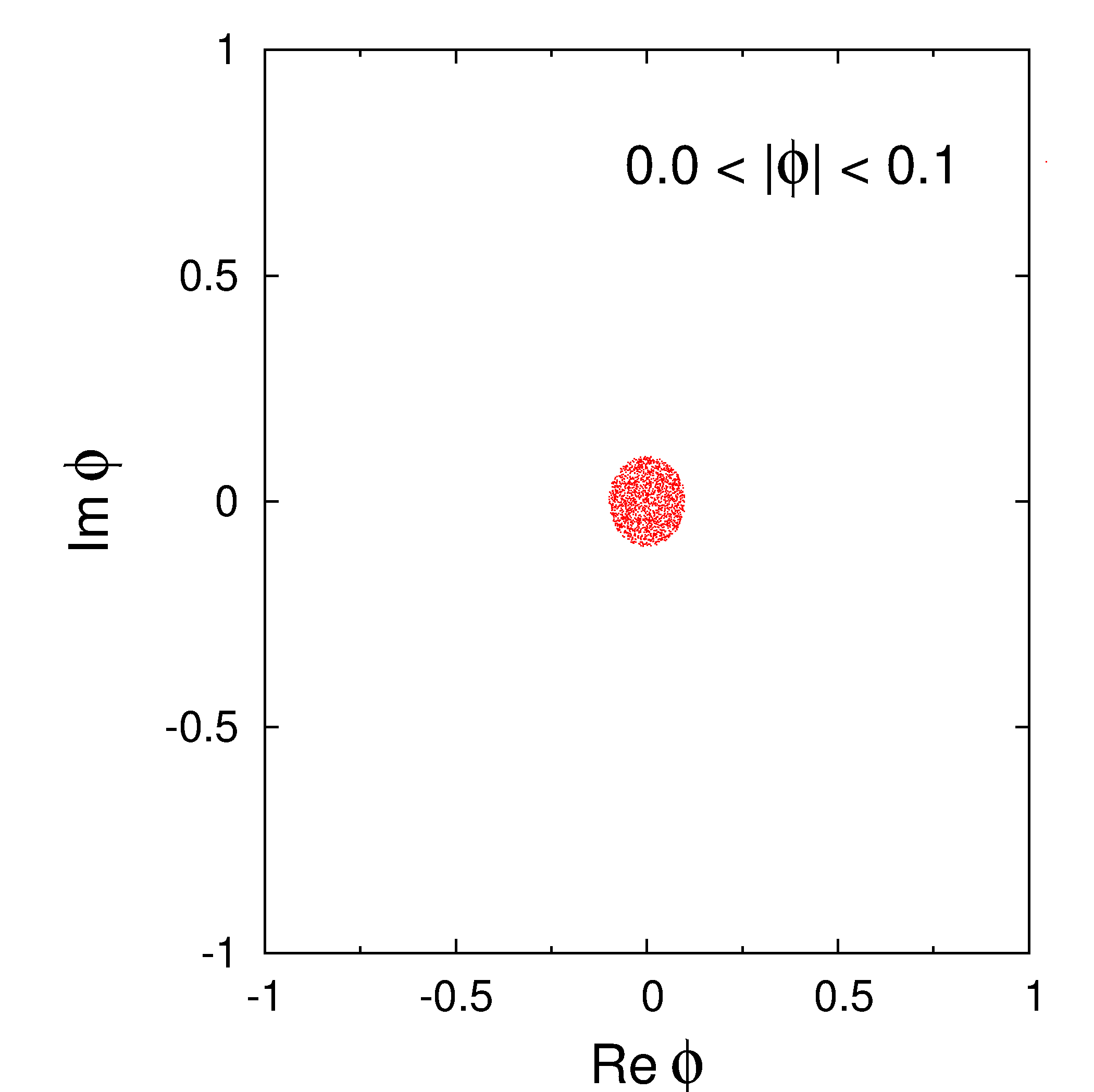}
\includegraphics[width=0.45\linewidth,height=0.3\linewidth, angle=0]{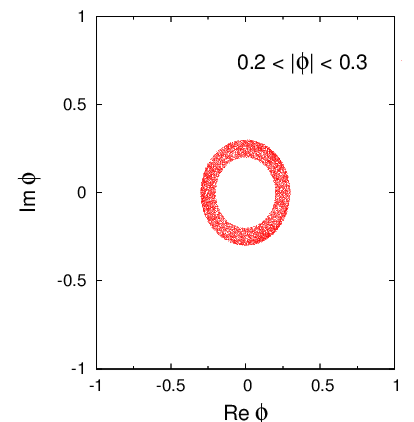}
\includegraphics[width=0.45\linewidth,height=0.3\linewidth, angle=0]{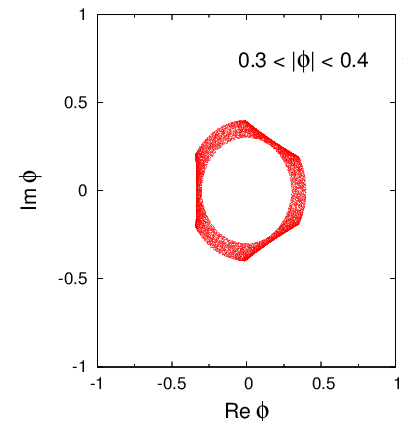}
\includegraphics[width=0.45\linewidth,height=0.3\linewidth, angle=0]{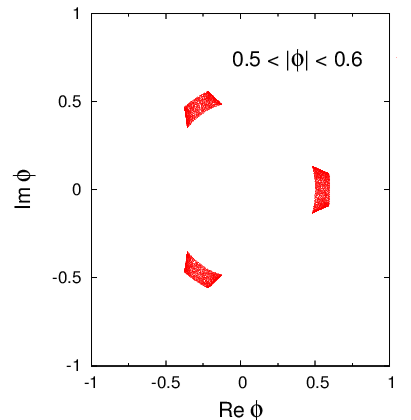}
\caption{(color online) The projection of $\Phi$ in Argand plane  for a fixed value, 
which are noted in the respective plots. This shows a strong color correlation in the 
range $0\leq \Phi \leq 1/3$ and ionization of color states in the range $1/3< \Phi \leq 1$.}
\label{phi_f_comp}
\end{center}
\end{figure}

All those features of center symmetry discussed above are also reflected in complex $\Phi$ plane in Fig.~\ref{phi_f_comp} indicating  
a snap shot of color localized and ionized domains of $Z(3)$. This clearly shows that there are center domains formation in the two
distinct regions of $|\Phi|$, which are  $0< |\Phi| \leq 1/3$ (confining domain) due to the center symmetry $Z(3)$, and  
$1/3 < |\Phi| \leq 1$ (deconfining or ionization domain) due to spontaneously breaking\footnote{Though also the $Z(3)$ symmetry is
explicitly broken with dynamical quarks unlike pure gauge sector, yet it can be regarded as an approximate symmetry and the Polyakov Loop 
expectation value is still useful as an order parameter as we will see later.}
 of the center symmetry $Z(3)$. 
The broken center domains formed in deconfining phases ($\langle L \rangle\ne 0$ or $1/3 < |\Phi| \leq 1$ )
will be separated by domain walls as they are distinguished by the phases $2\pi j/3$ with $j=0,1,2$.
Nevertheless, the formation of the center domains, the path of ionization and the distribution of the color charges from confining 
phase to deconfining phase or vice-versa will also depend on the nature of the color singlet potential for pure gauge (where $Z(3)$ is 
spontaneously broken) and also that with matter field (where $Z(3)$ is explicitly broken), which we discuss below using (\ref{potl}).

\subsection{Gauge Sector:}
The color singlet gauge potential is obtained in (\ref{potl}) as
\begin{eqnarray}
 \Omega^g_{S}&=&\Omega_g -\kappa T\ln H \nonumber \\
& =& 2  T \int\frac{d^3p}{(2\pi)^3} \ln
\left( 1 + \sum_{m=1}^8 a_m\, e^{-m\beta\epsilon_g}\right ) -\kappa T\ln H , \label{gpotl}  
\end{eqnarray}
where the first term describes the interaction of the spatial gluons with the Polyakov Loop 
(background temporal gauge field $A_0$ in (\ref{thfluc})) at finite $T$. 
The second term known as VdM term comes from invariant Haar measure.
The $Z(3)$ domains are plentiful (viz., Eq.(\ref{eq11})) as the gluon dynamics are 
solely governed by the thermal fluctuation of the background gauge field $A_0$  in (\ref{thfluc}).

\begin{figure}
\begin{center}
\includegraphics[width=0.45\linewidth,height=0.4\linewidth, angle=0]
{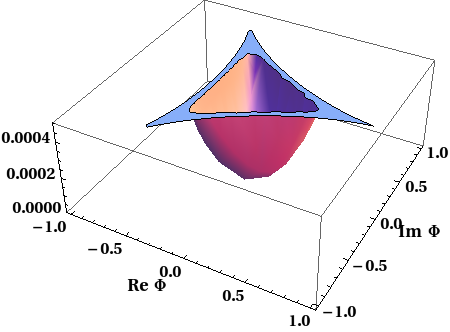}
\includegraphics[width=0.4\linewidth,height=0.35\linewidth, angle=0]
{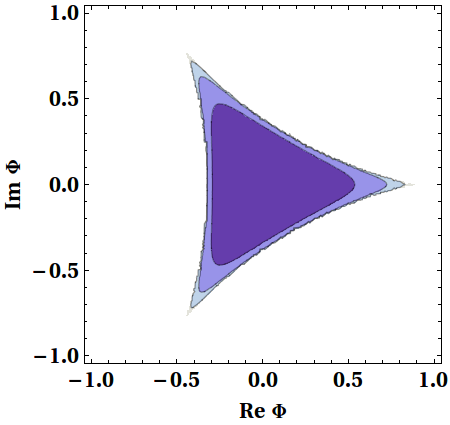}
\includegraphics[width=0.45\linewidth,height=0.4\linewidth, angle=0]
{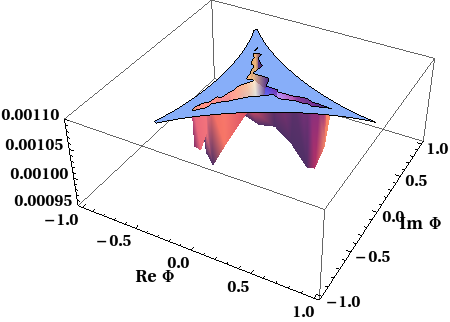}
\includegraphics[width=0.4\linewidth,height=0.35\linewidth, angle=0]
{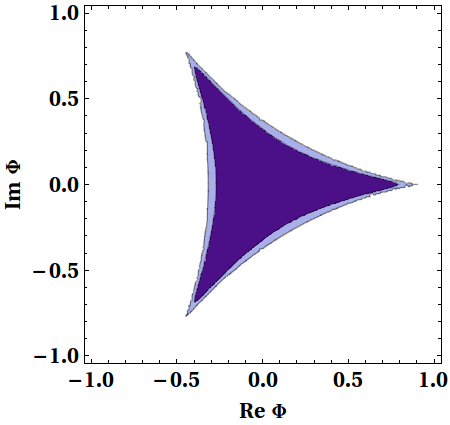}
\includegraphics[width=0.45\linewidth,height=0.4\linewidth, angle=0]
{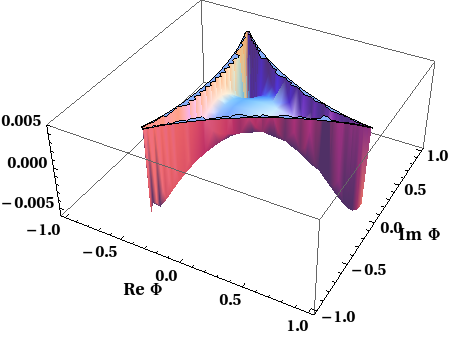}
\includegraphics[width=0.4\linewidth,height=0.35\linewidth, angle=0]
{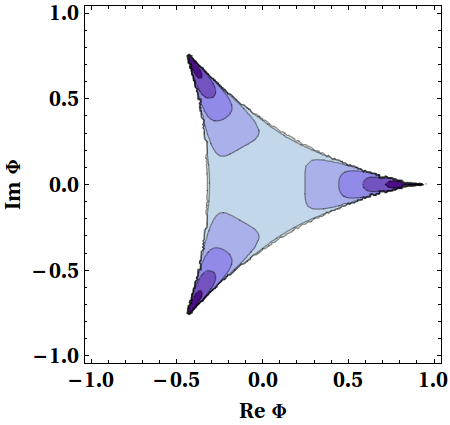}
\caption{(color online) {\it Left Panel:} A 3D plot of $\Omega_g\,  - \, \kappa T\ln \Theta_H$ in a 
complex $\Phi$ plane for $T=50,\, 270, \, {\mbox{and}} \,
350$ MeV and $\kappa=0.0075$ GeV$^3$.  {\it Right Panel:} Corresponding contour plots.}
\label{gauge_pot_fig}
\end{center}
\end{figure}

In Fig.~\ref{gauge_pot_fig} the color singlet gauge potential in a complex $\Phi$ plane is displayed for three temperatures. 
The left panel corresponds to 3-dimensional plots whereas right panel represents corresponding contour plots.
As can be seen the gauge potential, below $T< 270$ MeV,  has only one global minimum whereas for $T\ge 270$ MeV, it shows 
three minima representing a spontaneously broken $Z(3)$ phase for pure gauge.  The corresponding contour plots also display
the same features.  So, $T < 270$ MeV is color confining phase, where $Z(3)$ is unbroken as there is no ionization of
color charges. On the other hand $T\ge 270$ MeV there are ionization of color charges as the center symmetry is spontaneously broken
and those charges reside at the three minima in the potential in Fig.~\ref{gauge_pot_fig}  or three maxima in 
color space in Figs.~\ref{phi_3d_f} and \ref{phi_eqva} separated by distinct phases $2\pi j/3$ with $j=0,1,2$ and also  
by domain walls~\cite{bass}.  $T\sim (265-270)$ MeV, 
possibly indicates a phase transition for pure gauge and is in agreement with lattice result~\cite{boyd}.

In asymptotically high temperatures ($T\gg T_c$),  $\Phi,{\bar\Phi} \to 1$, $\langle {A}_0^2\rangle \to 0$, one recovers 
free gluon gas from $\Omega_g^{\rm{PL}}$ as 
\begin{eqnarray}
\Omega_S^{g; \; \Phi,{\bar\Phi}\rightarrow 1}
\!\!\! &=& \!\! 2 (N_c^2-1) T \! \! \int\frac{d^3p}{(2\pi)^3}
\ln\left( 1 - e^{-\beta \epsilon_g} \right),
\label{eq12}
\end{eqnarray}
where $N_c=3$. The VdM term due to the invariant Haar measure disappears and the spatial gluons are completely ionized.  

\begin{figure}
\begin{center}
\includegraphics[width=0.35\linewidth,height=0.25\linewidth, angle=0]{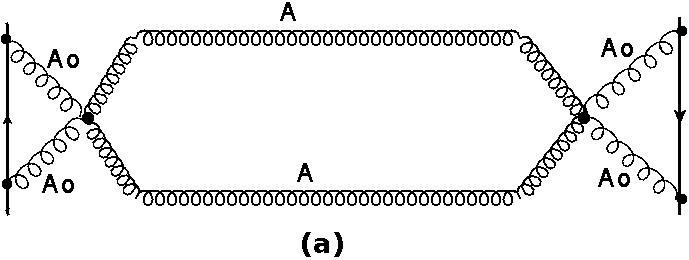}
\,\,\,\, \,\,\,\, \includegraphics[width=0.35\linewidth,height=0.25\linewidth, angle=0]{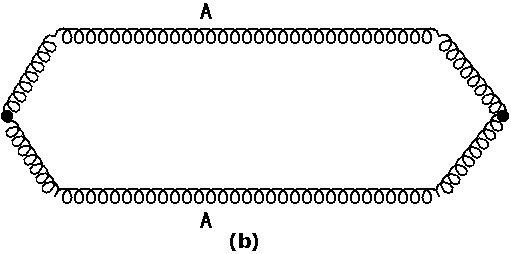}
\caption{(a) An exchange of a pair of massive magnetic gluons between two Polyakov Loops (b) Spontaneously created gluon 
condensate when $\Phi=0$.}
\label{mag}
\end{center}
\end{figure}

At low temperature ($T\ll T_c$),  the amplitude of $\langle A_0^2\rangle$ is high and $\Phi\rightarrow 0$, 
$\Omega_{S}^g$ becomes
\begin{eqnarray}
\Omega_S^{g;\;\Phi,{\bar\Phi}\rightarrow 0} 
= 2T \int\frac{d^3p}{(2\pi)^3} \left [
\ln\left( 1 - e^{-N_c\beta \epsilon_g} \right)^{2}\, \right.  
&& \nonumber \\ 
\left. \! +\! \ln\left( 1 - e^{2\pi i/N_c}e^{-\beta\epsilon_g} \right)
\! +\! \ln\left( 1 - e^{-2\pi i/N_c}e^{-\beta\epsilon_g} \right)\right ],
\label{eq13}
\end{eqnarray}
where the color charges are frozen through color-singlet states in the confining domain.
Now (\ref{eq13}) can be viewed in the following way: 

(i) The first term 
indicates that the  Polyakov Loop confines $N_c$ number of spatial gluons in a same energy 
state representing a glueball. There are two such copies which is consistent 
with $SU(3)$ gauge theory as $8\otimes 8\otimes 8$ generates only two singlet 
glueball states. Obviously, the $Z(3)$ charge is frozen in the color singlet 
glueballs through the localization of $Z(3)$ charge in the global $Z(3)$ 
minimum in Fig.~\ref{gauge_pot_fig}. The first term is negative which generates positive
pressure in QCD confined object.

(ii) The remaining two spatial gluons in the second and third terms
are conjugate to each other but distinguished by $Z(3)$ phase.
In a confined phase ($T\ll T_c; \ \Phi\sim 0$), Polyakov Loop disappears but this two spatial gluons cannot
stay as  free color charges rather they  keep circulating in a  closed loop of virtual color and anti-color charges 
as shown in the right panel of Fig.~\ref{mag} in a nonperturbative 
vacuum. The potential combining this two spatial gluons with $Z(3)$ phases
can then be written as
\begin{eqnarray}
&& 2T \int\frac{d^3p}{(2\pi)^3}
\ln\left[ \left( 1 - e^{2\pi i/N_c}e^{-\beta\epsilon_g} \right)
\left( 1 - e^{-2\pi i/N_c}e^{-\beta\epsilon_g} \right)\right ] \nonumber \\
\nonumber \\ 
&& = 2T \int\frac{d^3p}{(2\pi)^3}
\ln\left( 1 + e^{-\beta \epsilon_g}+e^{-2\beta\epsilon_g} \right)\,,
\label{eq13c}
\end{eqnarray}
which is a positive definite quantity. Such states, may be condensates, are produced spontaneously in a nonabelian 
gauge theory as the pressure generated by this two spatial gluons in terms of the center group $Z(3)$ is negative 
compared to the confined object (first term in (\ref{eq13})). This two  gluons should not contribute directly 
to the thermodynamics in a confined phase. Rather they provide a nonperturbative ground state pressure which is negative and unbound 
from below that can be viewed as a general confining background of strong interaction. In lattice QCD calculations of equation of states 
this confining background is removed so that the pressure starts from zero or positive value in the confined phase (first term in (\ref{eq13})), 
i.e., in the  low temperature phase ($T\ll T_C$). 
    
It is also worth noting here that in Ref.~\cite{arnold} the two conjugate spatial gluons has been considered 
in the nonzero $\Phi$ domain as an exchange of a pair of massive spatial virtual gluons between two Polyakov Loops as 
shown in the left panel of Fig.~\ref{mag}. This is because Euclidean time
reflection does not allow an exchange of a massless pair of spatial gluons~\cite{arnold}.
This was shown to generate a magnetic screening mass ($\sim g^2T$), a solely non-perturbation correction 
to the electric screening mass. This in turn provides a nonperturbative mass gap 
to prevent those excitations of color charges having energy lower than it. In the confining phase ( $|\Phi|=0$), the Polyakov Loop
disappears and this two magnetic gluons keep circulating  in a closed loop that provides a confining background as discussed above. 


\begin{figure}
\begin{center}
\includegraphics[width=0.45\linewidth,height=0.3\linewidth, angle=0]
{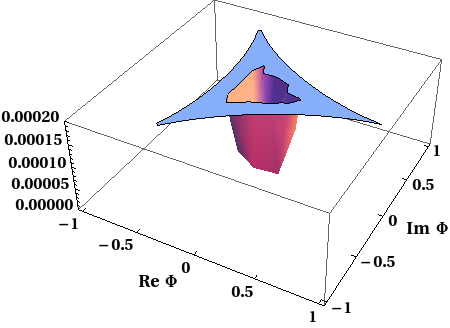}
\includegraphics[width=0.4\linewidth,height=0.3\linewidth, angle=0]
{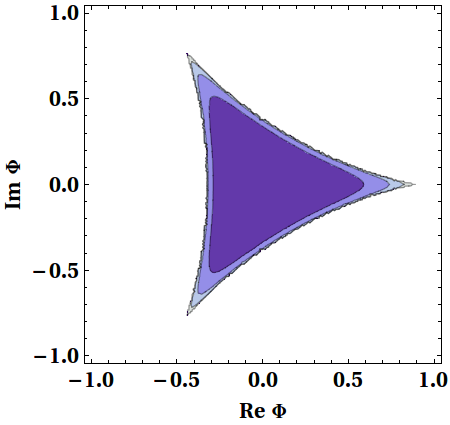}
\includegraphics[width=0.45\linewidth,height=0.3\linewidth, angle=0]
{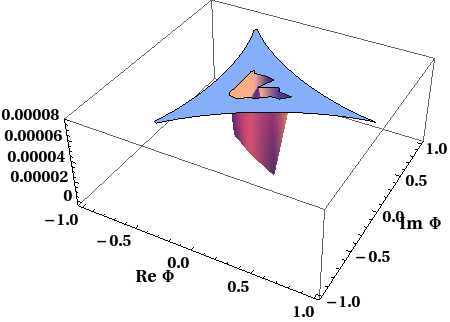}
\includegraphics[width=0.4\linewidth,height=0.3\linewidth, angle=0]
{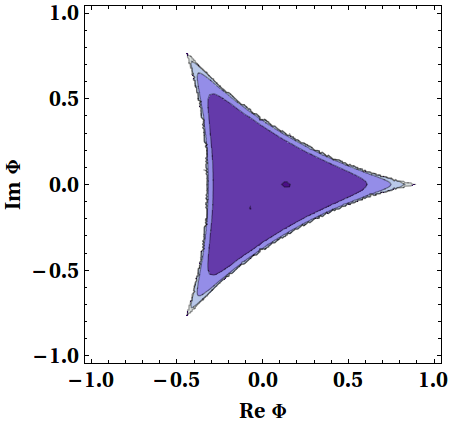}
\includegraphics[width=0.45\linewidth,height=0.3\linewidth, angle=0]
{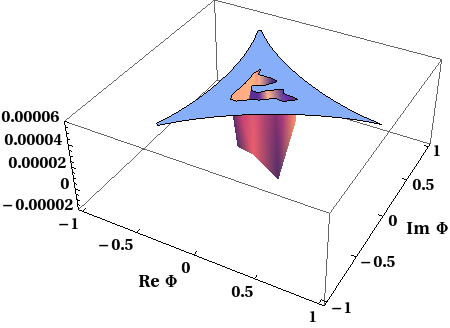}
\includegraphics[width=0.4\linewidth,height=0.3\linewidth, angle=0]
{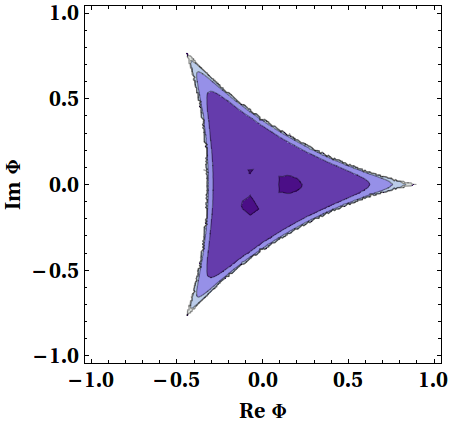}
\includegraphics[width=0.45\linewidth,height=0.3\linewidth, angle=0]
{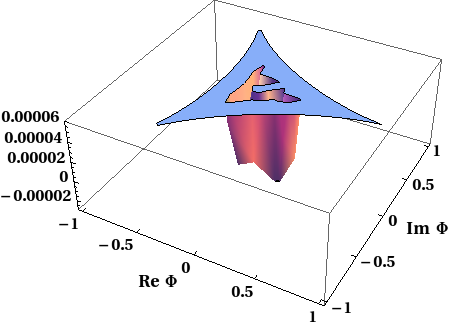}
\includegraphics[width=0.4\linewidth,height=0.3\linewidth, angle=0]
{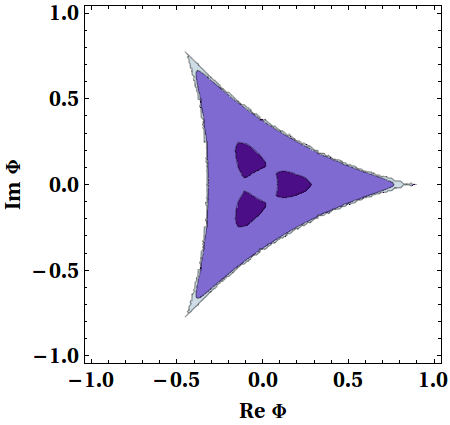}
\includegraphics[width=0.45\linewidth,height=0.3\linewidth, angle=0]
{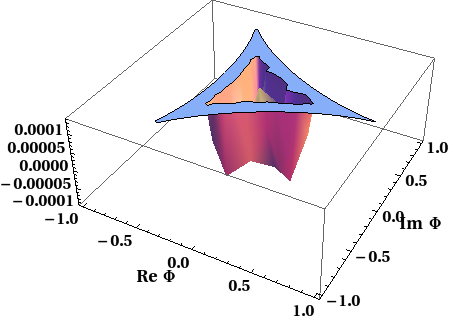}
\includegraphics[width=0.4\linewidth,height=0.3\linewidth, angle=0]
{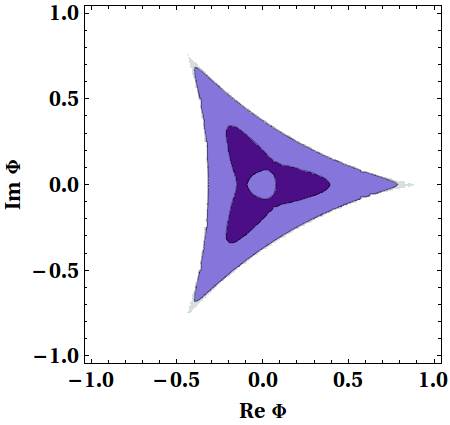}
\end{center}
\end{figure}

\begin{figure}
\begin{center}
\includegraphics[width=0.4\linewidth,height=0.3\linewidth, angle=0]
{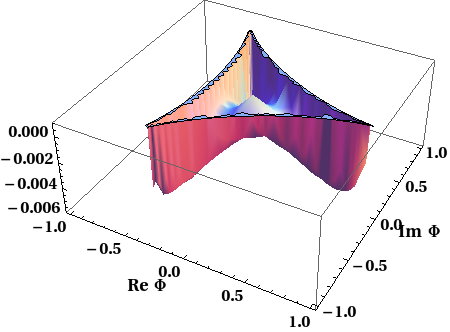}
\includegraphics[width=0.4\linewidth,height=0.3\linewidth, angle=0]
{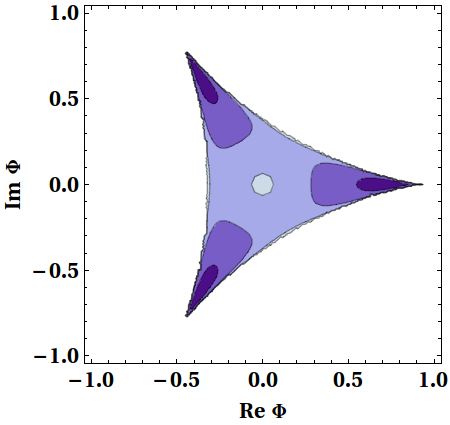}
\caption{(color online) {\it Left Panel:} A 3D plot of full potential $\Omega=\Omega_q+\Omega_{\bar q}+\Omega_g\,  - \, 
\kappa T \ln H$ in a complex $\Phi$ plane for $T=140,\, 149,\, 155, \, 160, \, 170 \, \, \, {\mbox{and}} \, \, \, 
250$ MeV with $\mu=0$ and $\kappa=0.0075$ GeV$^3$. 
{\it Right Panel:} Corresponding contour plots.}
\label{full_pot_fig}
\end{center}
\end{figure}

\subsection{Matter sector:} The full color singlet potential in presence of matter field is given by (\ref{potl}) as
\begin{eqnarray}
\Omega^{S}=\Omega_q+\Omega_{\bar q}+\Omega_g-\kappa T\ln H . \label{mpotl}
\end{eqnarray}
The $Z(3)$ symmetry with quarks and antiquarks in the full potential is explicitly broken  
under the rotation of $Z(3)$ since they also carry the $Z(3)$ charge. This can be viewed from Fig.~\ref{full_pot_fig} that 
displays a plot of $\Omega^{S}$ in a complex $\Phi$ plane for a set of temperatures. The left panel in Fig.~\ref{full_pot_fig}
corresponds to 3-dimensional plots whereas right panel corresponds to contour plots in a complex $\Phi$ plane. 
In the range $ 0 <T < 140$ MeV, the potential has only one minimum (not shown here in Fig.~\ref{full_pot_fig}) that apparently 
represents $Z(3)$ global minimum. This is
because the  $Z(3)$ color charges are still frozen in hadrons in the confining domain. 
In the temperature domain $140 < T (\rm{MeV}) < 150$, there is still one minimum but has moved away from the $Z(3)$ center 
and does not remain symmetric under $Z(3)$ rotation $2\pi/3$. Here, the mean freepath of the color charges is determined by
the effective size of the $Z(3)$ domains~\cite{bass}.  For $150 \le T(\rm{MeV}) < 170$,  
the potential shows one global minimum with larger depth and two local minima with smaller but uneven depth.  This implies that 
all three minima are not symmetric under the  $Z(3)$ rotation $2\pi/3$. This is unlike the pure gauge case in Fig.~\ref{gauge_pot_fig}, 
an indication of explicit $Z(3)$  symmetry breaking in presence of dynamical quarks. In this region $150 \le T(\rm{MeV}) < 170$ the mean freepath 
of the color charges begin to increase as they tend to move from local minima to energetically favourable global minimum.  
 
Interestingly, the explicit symmetry breaking due to the presence of the matter fields leads to a {\it metastable state} in the temperature range 
($145\le T (\rm{MeV}) \le 170$) and beyond which the system crosses smoothly to
the deconfined phase. So, the concept of $T_c$  is not very well defined and lattice QCD calculations estimate it through 
various observable~\footnote{The analysis of HISQ/{\it tree} and {\it asqtad} 
action by HotQCD collaboration give a consistent results~\cite{peter} in the continuum 
limit $T_c=(159\pm9)$ MeV from the peak of susceptibilites. The Wuppertal-Budapest collaboration~\cite{wb} using {\it stout} action 
found $T_c=147(2)(3)$ MeV, $157(3)(3)$ MeV and $155(3)(3)$ MeV from  the peak position of susceptibility, and 
inflection points in chiral and renormalized chiral condensates.} and find in the range $(147-160)$ MeV. 
Our observation of $T_c\sim (165- 170)$ MeV where the instability of the metastable minima stabilizes due to 
the expansion of the domains as can be seen clearly from the contour plots in the right panel of Fig.~\ref{full_pot_fig}. 
Like gauge sector the depth of all three minima becomes almost symmetric for $T\ge 170 $MeV 
as the domains expand with temperature. The color charges, irrespective of their nature, begin to reside at those  minima in 
Fig.~\ref{full_pot_fig} for $T\ge 170$ MeV.
These domains are separated by non-perturbative domain walls even well above $T_c$ because the fluctuations of 
the background gauge field are still nonzero.  Moreover, the mean free path of the color charges becomes of the order of 
the effective size of these domains~\cite{bass}.  Since the domains expand, arround $T\ge T_c$ the domain size is usually smaller 
that corresponds to shorter wavelength appropriate for hydrodynamics to be applicable
whereas the perturbative QCD may be applicable at high $T$ as the domain size increases. There will also be plenty of  domains
at high $T$ due to the fluctuations of the background gauge field.
In the color deconfined phase these domains as well domain walls act as scattering centers that cause high energy jet 
to lose energy through gluon radiations and get quenched~\cite{bass}. 

In the asymptotically high temperature ($T\gg T_c$),  eq.(\ref{mpotl}) becomes
\begin{eqnarray}
\Omega_S^{\Phi,{\bar\Phi}\rightarrow 1}  
\!\!\! &=& \!\! -2N_fT \int\frac{d^3p}{(2\pi)^3}
\ln\left( 1 + e^{-\beta(\epsilon_q-\mu)} \right)^{N_c} \nonumber \\
&& -2N_fT \int\frac{d^3p}{(2\pi)^3}
\ln\left( 1 + e^{-\beta(\epsilon_q+\mu)} \right)^{N_c} \nonumber \\
&& + \Omega_g^{{\rm{PL}};\Phi,{\bar\Phi}\rightarrow 1}\, , 
\label{eq7}
\end{eqnarray}
where $N_c=3$. It represents the thermodynamic potential for
a free colored quark, antiquark and gluon gas at high temperature, i.e, the color charges are
completely ionized and reside at those minima in potential of Fig.~\ref{full_pot_fig} or equivalently at those 
maxima in color space of Figs.~\ref{phi_3d_f} and \ref{phi_eqva}. 

At low temperature ($T\ll T_c$), the potential with matter part can be written as
\begin{eqnarray}
\Omega^{\rm{PL};\Phi,{\bar\Phi}\rightarrow 0} 
&=&\!\! -2N_fT \!\! \int \!\! \frac{d^3p}{(2\pi)^3}
\ln\left( 1 + e^{-\beta N_c(\epsilon_q-\mu)} \right) \nonumber \\
&& -2N_fT \!\! \int \!\! \frac{d^3p}{(2\pi)^3}
\ln\left( 1 + e^{-\beta N_c(\epsilon_q+\mu)} \right) \nonumber \\
&&+ \Omega_g^{{\rm{PL}};\Phi,{\bar\Phi}\rightarrow 0} 
\,.
\label{eq8}
\end{eqnarray}
This represents the thermodynamic potential for a composite color singlet
object containing three quarks(antiquarks) in a same color state with the same energy. So 
is for gluons as discussed earlier. One can also combine appropriate terms in (\ref {eq8}) to get mesons, baryons, 
hybrid mesons, glueballs etc.
In other words, the $SU(3)$ color singlet restriction vis-a-vis Polyakov Loop  dynamically confines three 
colored charges in a same energy state, which finally forms a color neutral
object like  baryon and glueball. This is because color charges get frozen in color singlet states like hadrons 
in the global minimum when $T\ll T_c$ . 
This essentially boils down to the fact that the $SU(3)$ color singlet restriction vis-a-vis Polyakov Loop dynamically 
provides the basis for the recombination of partons for hadronisation
from quark-gluon plasma when it cools down  below $T_c$. In Refs.~\cite{mus2} the colorsingletness has explicitly 
shown to provide the natural explanation of the scaling law (of the valence partons) of the eliptic flow of the identified 
hadrons in heavy-ion collisions, a direct evidence of deconfined phase~\cite{rhic,hadr}.

\section{Conclusion:} We show that the color singlet ensemble of a quark-gluon gas  becomes equivalent to that
of Polyakov Loop Model within a stationary point approximation. The calculation is based on quantum statistical mechanics 
with a global $SU(3)$ symmetry but considering the Haar measure at each spatial points to take into account the 
confinement effect. The normalized character in fundamental representation of $SU(3)$ exhibits center symmetry, 
$Z(3)$, of $SU(3)$ akin to Polyakov Loop. In the process, we have also obtained pure gauge potential explicitly. 

The color singlet gauge potential shows center symmetry which is spontaneously broken in high temperature phase ($T\ge 270$ MeV).  
When matter field is added the center symmetry is found to be broken explicitly, which leads to a metastable state 
in the temperature domain $145 \le T(\rm{MeV}) \le 170$. The instability of the metastable state stabilizes for $T\ge 170$MeV and
there are domains formed in the deconfined phase. We also discussed the phenomenological consequences of these center domains, both in
pure gauge as well as with dynamical quarks, on color confining-deconfining phase transition or vice-versa in QCD, through the color 
singlet vis-a-vis Polyakov Loop potential. The center symmetry dictates that the confined phase 
appears as a color singlet object from the dynamical recombination of three partons as given in (\ref{eq8}), plus a 
confining background. This would solely describe the thermodynamic properties of color singlet structures like baryon, 
antibaryon, meson and glueball. Most of the effects of heavy-ion collisions: non-perturbative nature 
of the deconfined phase, fluid nature, jet quenching, recombination of hadronization etc can be understood in terms of the center domains.
More calculations in this direction are required to make quantitative predictions on the consequences of center domains in heavy-ion 
phenomenology. 

\vspace*{0.3in}

\noindent{\it{Acknowledgment:}} CAI would like to acknowledge the financial support from University Grants Commission.

\end{document}